\documentstyle[aps,epsfig,floats,amssymb]{revtex}
\draft
\newcommand{\om}{\omega}

\newcommand{\txt}{\textstyle}

\newcommand{\beq}{\begin{equation}}
\newcommand{\eeq}{\end{equation}}
\newcommand{\ba}{\begin{array}}
\newcommand{\bea}{\begin{eqnarray}}
\newcommand{\ea}{\end{array}}
\newcommand{\eea}{\end{eqnarray}}

\newcommand\comment[1]{ \hbox{[{\it Comment suppressed here.}\/]} }
\newcommand\hide[1]{}

\newcommand{\skipover}[1]{}
\newcommand{\nn}{\nonumber}
\newcommand{\half} {{\txt {1\over 2}}}

\newcommand{\Tr}{\hbox{Tr}}

\newcommand{\vecz}{{\mathbf z}}

\newcommand{\be}{\begin{equation}}
\newcommand{\ee}{\end{equation}}
\newcommand{\bean}{\begin{eqnarray*}}
\newcommand{\eean}{\end{eqnarray*}}

\newcommand{\rmd}{{\rm d}}

\newcommand{\C}{{\cal C}}

\begin{document}                                                

%
\twocolumn[\hsize\textwidth\columnwidth\hsize\csname
@twocolumnfalse\endcsname

\title{
Classical aspects of quantum fields far from equilibrium 
}

\author{Gert Aarts and J\"urgen Berges}

\address{
Institut f{\"u}r Theoretische Physik,
Philosophenweg 16, 69120 Heidelberg, Germany}

\date{July 11, 2001}

\maketitle

\begin{abstract} 
We consider the time evolution of nonequilibrium quantum scalar fields in
the $O(N)$ model, using the next-to-leading order $1/N$ expansion
of the $2PI$ effective action. A comparison with exact numerical
simulations in $1+1$ dimensions in the classical limit
shows that the $1/N$ expansion gives quantitatively precise results
already for moderate values of $N$. For sufficiently high initial 
occupation numbers the time evolution of quantum fields is shown to 
be accurately described by classical physics. Eventually the correspondence
breaks down due to the difference between classical and quantum thermal
equilibrium.
\end{abstract}
\pacs{PACS numbers: 11.15.Pg, 11.10.Wx, 05.70.Ln  \hfill HP-THEP-01-28}
%
 ]
   

In recent years we have witnessed an enormous increase of interest 
in the dynamics of quantum fields out of equilibrium. Much progress has
been achieved for systems close to thermal equilibrium or with 
effective descriptions based on a separation of scales in the
weak coupling limit \cite{Bodeker:2001pa}. Current 
and upcoming relativistic heavy-ion collision experiments provide
an important motivation to find controlled nonperturbative
approximation schemes which yield a quantitative description of 
far-from-equilibrium phenomena from first principles. 

Practicable nonperturbative approximations may be based on the
two-particle irreducible ($2PI$) 
generating functional for Green's functions \cite{Cornwall:1974vz}.
Recently, a systematic $1/N$ expansion of the $2PI$ effective action
has been proposed for a scalar $O(N)$ symmetric quantum field theory
\cite{Berges:2001fi}. This nonperturbative approach 
extends previous successful descriptions of the large-time 
behavior of quantum fields \cite{Berges:2000ur,Aarts:2001qa}, which employ
the loop expansion of the $2PI$ effective action relevant
at weak couplings \cite{Cornwall:1974vz,Calzetta:1988cq}. 
At next-to-leading order (NLO) the $1/N$ expansion of the $2PI$ effective 
action has been solved for the quantum theory in $1+1$ 
dimensions \cite{Berges:2001fi}. The approach  
overcomes the problem of a secular time evolution, which is encountered 
in the standard $1/N$ expansion of the $1PI$ effective action beyond 
leading order \cite{Mihaila:2000ib}. 

Our purpose in this Letter is to establish that the $1/N$ expansion at NLO
gives quantitatively precise results already for moderate values of $N$.
We therefore have a small nonperturbative expansion parameter at hand
and a controlled description of far-from-equilibrium dynamics becomes 
possible. As an
application we compare quantum and classical evolution and demonstrate
that the nonequilibrium quantum field theory can be described by its
classical field theory limit for sufficiently high initial occupation
numbers. We show that eventually the correspondence breaks down due to the
difference between classical and quantum thermal equilibrium.

Employing the classical statistical field theory limit, we compare the
$1/N$ expansion of the $2PI$ effective action at NLO with the {\em
exact}\, result, which includes all orders in $1/N$. 
The time evolution of classical nonequilibrium Green's functions can be
calculated exactly, up to controlled numerical uncertainties, by
integrating the microscopic field equations of motion.
This allows one to obtain a direct comparison \cite{Aarts:2001wi}.  
Apart from benchmarking approximation schemes employed in quantum field
theory, the importance of the classical field limit for the approximate
description of nonequilibrium quantum fields is manifest.

{\em $2PI$ effective action.}
We consider a real \mbox{$N$-component}
scalar quantum field theory with a $\lambda
(\phi_a\phi_a)^2/(4! N)$ interaction in the symmetric
phase ($a=1,\ldots,N$). The $2PI$ generating functional for Green's 
functions can be parametrized as \cite{Cornwall:1974vz}
\be
\Gamma[G] = \frac{i}{2} \Tr\ln G^{-1} 
          + \frac{i}{2} \Tr\, G_0^{-1} G
          + \Gamma_2[G] + {\rm const},  
\label{2PIaction}
\ee 
where $G_0^{-1}=i(\square+m^2)$ denotes the free inverse propagator.  
The $2PI$ contribution $\Gamma_2[G]$ can be computed from a systematic 
$1/N$ expansion of the $2PI$ effective \mbox{action \cite{Berges:2001fi}}. 
Writing  
$
\Gamma_2[G]= \Gamma_2^{\rm LO}[G] + \Gamma_2^{\rm NLO}[G] +\ldots
$
the LO and NLO contributions are given by
\cite{Berges:2001fi,Mihaila:2001sr}
\bea
&&\Gamma_2^{\rm LO}[G] = - \frac{\lambda}{4! N} 
  \int_{\C} {\rmd}^{d+1}x\, G_{aa}(x,x) G_{bb}(x,x), 
\label{LOcont} 
\\
&&\Gamma_2^{\rm NLO}[G] =  \frac{i}{2}\, \Tr_{\C}
\ln [\, {\bf B}(G)\, ] \, .
\label{NLOcont} 
\eea
Here $\C$ denotes the Schwinger-Keldysh contour along the real
time axis \cite{Schwinger:1961qe}
and 
\beq
{\bf B}(x,y;G) = \delta_{\C}^{d+1}(x-y)
+ i \frac{\lambda}{6 N}\, G_{ab}(x,y)G_{ab}(x,y).
\label{Vertex}
\eeq
The nonlocal four-point vertex at NLO is given by  
$\frac{\lambda}{6N}\, {\bf B}^{-1}$ \cite{Berges:2001fi}.
In absence of external sources the evolution equation for $G$ is
determined by \cite{Cornwall:1974vz}
\beq
\frac{\delta \Gamma[G]}{\delta G_{ab}(x,y)} = 0.
\label{stationary}
\eeq
We consider the $O(N)$ symmetric case, such that $G_{ab}(x,y) =
G(x,y)\,\delta_{ab}$. In the following we solve Eq.~(\ref{stationary}),
using Eqs.~(\ref{2PIaction})--(\ref{Vertex})
without further approximations numerically in $1+1$ 
dimensions \cite{Berges:2001fi}. We compare
the outcome with results in the classical field theory limit, using both
the NLO classical approximation and the ``exact'' Monte Carlo calculation
\cite{Aarts:2001wi} as described below.

{\em Nonequilibrium time evolution.}
To formulate the non\-equilibrium dynamics as an initial-value problem, 
we decompose the full two-point function using the identity 
$
G(x,y) = F(x,y)- (i/2) \rho(x,y)\, {\rm sign}_{\C}(x^0-y^0)
$
where $F$ is the symmetric or statistical two-point function and $\rho$
denotes the spectral function
\cite{Aarts:2001qa}. Following Refs.\ \cite{Aarts:2001qa,Berges:2001fi},
Eq.\ (\ref{stationary}) can be written as
\bea
\nn
\left[\square_x +M^2(x)\right]F(x,y)\! &=&
- \int_0^{x^0}\!\!\! \rmd z^0
\!\!\int \! \rmd\vecz\,\, \Sigma_{\rho}(x,z)F(z,y) \nonumber \\
&+& \int_0^{y^0}\!\!\! \rmd z^0 \!\!\int \!\rmd\vecz\,\, \Sigma_{F}(x,z)
\rho(z,y), 
\label{eqF1}
\\ 
\left[\square_x +M^2(x)\right]\rho(x,y) &=&
-\int_{y^0}^{x^0}\!\!\! \rmd z^0 \!\!\int \! \rmd\vecz\,\,
\Sigma_{\rho}(x,z)\rho(z,y).
\nonumber
\eea
At NLO in the $1/N$ expansion the effective mass term $M^2(x)$ is given by 
$
M^2(x) = m^2 +\lambda\frac{N+2}{6N}F(x,x)
$
and the self energies are \cite{Berges:2001fi}
\bea
\Sigma_{F}(x,y) &=\!& -\frac{\lambda}{3N} \Big[ F(x,y)I_{F}(x,y)
- \frac{1}{4} \rho(x,y) I_\rho(x,y) \Big], \!\!\!
\label{sigmaF}\\
\Sigma_{\rho}(x,y) &=\!& -\frac{\lambda}{3N}
\Big[\rho(x,y)I_{F}(x,y)+F(x,y)I_{\rho}(x,y)\Big]. \!\!\!
\label{sigmarho}
\eea
Here the functions $I_{F}$ and $I_{\rho}$ resum an infinite chain of bubble 
diagrams, 
\bean
I_{F}(x,y) = 
-\frac{\lambda}{3}\Pi_{F}(x,y)
+\frac{\lambda}{3}
\int_0^{x^0}\!\!\! \rmd z^0
\int \! \rmd\vecz\,\, I_{\rho}(x,z)\Pi_{F}(z,y)  
&&\nonumber \\
-\frac{\lambda}{3}
\int_0^{y^0}\!\!\! \rmd z^0 \int \!\rmd\vecz\,\, I_{F}(x,z)\Pi_{\rho}(z,y),
&&\\
\nn
I_{\rho}(x,y) = 
-\frac{\lambda}{3}\Pi_{\rho}(x,y)
+\frac{\lambda}{3}
\int_{y^0}^{x^0}\!\!\! \rmd z^0 \int \! \rmd\vecz\,\,
I_{\rho}(x,z)\Pi_{\rho}(z,y),
&&
\eean
with
\bea
\Pi_{F}(x,y) &=& -\frac{1}{2}\Big(F^2(x,y)-\frac{1}{4}\rho^2(x,y)\Big),
\label{PIF}\\
\Pi_{\rho}(x,y) &=& -F(x,y)\rho(x,y).
\label{PIR}
\eea

{\em Classical field theory limit.} The classical statistical field theory
limit of a scalar quantum field theory has been studied extensively
in the literature. An analysis along the lines of Refs.\
\cite{Aarts:1997qi,Buchmuller:1997yw,Cooper:2001bd,Blagoev:2001ze}
shows that all equations (\ref{eqF1})--(\ref{PIR}) remain the same in the
classical limit except for
differing expressions for the statistical components of the self energy
\bea
\label{eqSigmacl}
\Sigma_{F}(x,y) & {\rm classical\, limit} \atop \Longrightarrow & 
-\frac{\lambda}{3N} F(x,y)I_{F}(x,y),\\
\Pi_{F}(x,y) &{\rm classical\, limit} \atop \Longrightarrow & 
-\frac{1}{2}F^2(x,y).
\label{eqPicl}
\eea
The latter expressions are given by Eqs.\ (\ref{sigmaF}) and (\ref{PIF})
in the quantum theory. One observes that the classical self energies are
obtained from the 
expressions in the quantum theory by dropping terms with two spectral
($\rho$-type) components compared to two statistical ($F$-type) functions. 
This particular relationship has been studied in great detail in thermal
equilibrium and corresponds to retaining only
contributions that are of leading order in $\hbar$. It has been
systematized in terms of Feynman rules for classical and quantum theories
using the Keldysh formulation with appropriate interaction vertices
\cite{Aarts:1997qi}. Note that the classical spectral
function is obtained from the quantum one by replacing $-i$ times the
commutator by the classical Poisson bracket. 
A comparison of the classical limit in the current
approximation for $N=1$ has been studied  in Ref.\ \cite{Blagoev:2001ze}.

{\em Monte Carlo approach.} 
An ``exact'' nonperturbative solution of the evolution of classical
correlation functions in the $O(N)$ model can be obtained numerically in a
straightforward manner \cite{Aarts:2001wi}.  Initial conditions are
determined from a probability functional on classical phase-space. The
subsequent time evolution is solved numerically using the classical
equations of motion.  In the figures presented below, we have sampled
50000-80000 independent initial conditions to approximate the exact
evolution.

{\em Far from equilibrium evolution.}
We consider a system that is invariant under space translations and work
in momentum space.
We choose a Gaussian initial state such that a specification of the
initial two-point functions is sufficient. 
The quantum (classical) spectral function at initial time is completely
determined from the equal-time commutation relations (Poisson brackets). 
For the symmetric two-point function we take $F(0,0;p) = [n_0(p)+\half]/\om_p$,
with the initial particle number $n_0(p)= n_{\rm ts}(p)+ n_B(p)$   
representing a peaked ``tsunami'' 
$
n_{\rm ts}(p) = {\cal A} \exp\left[ 
-\frac{1}{2 \sigma^2}(|p|-|p_{\rm ts}|)^2\right]
$
in a thermal background
$
n_B(p) = [\exp (\om_p/T_0) -1]^{-1}
$
\cite{Pisarski:1997cp,Aarts:2001qa,Berges:2001fi}. 
Such an initial state is reminiscent of two colliding wave packets
\cite{Pisarski:1997cp} and provides a far-from-equilibrium initial
condition. We emphasize that these initial conditions can be implemented
both in a quantum and a classical theory.
The initial mode energy is given by $\om_p = (p^2+M^2)^{1/2}$
where $M$ is the one-loop renormalized mass in presence of the
nonequilibrium medium, determined from the one-loop gap equation. 
As a renormalization condition we choose the one-loop renormalized mass in
vacuum $m_R\equiv M|_{n_0=0}=1$ as our dimensionful scale.
The results shown below are obtained using a fixed coupling constant 
$\lambda/m_R^2=30$. 

\begin{figure}
\begin{center}
\epsfig{file=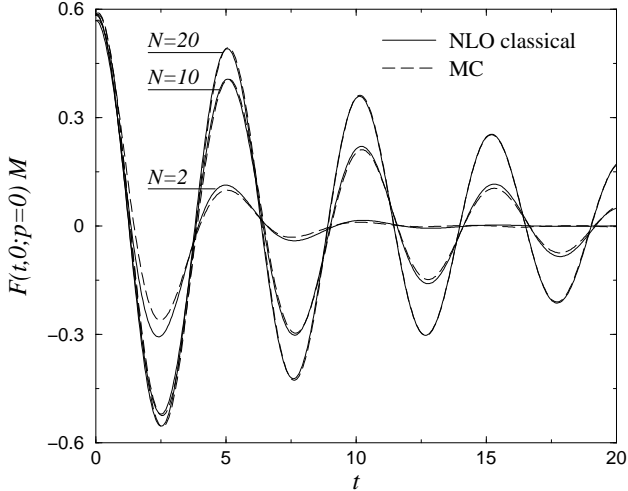,height=6.8cm}
\end{center}
\caption{Unequal time two-point function $F(t,0;p=0)$ at zero
momentum times the initial mass $M$ for $N=2,10,20$. The full lines show
results from the NLO classical evolution and the dashed lines from the
exact classical evolution (MC). For $N=20$ the NLO and exact evolution
can hardly be distinguished.
}
\label{fig1}
\end{figure}

\begin{figure}
\begin{center}
\epsfig{file=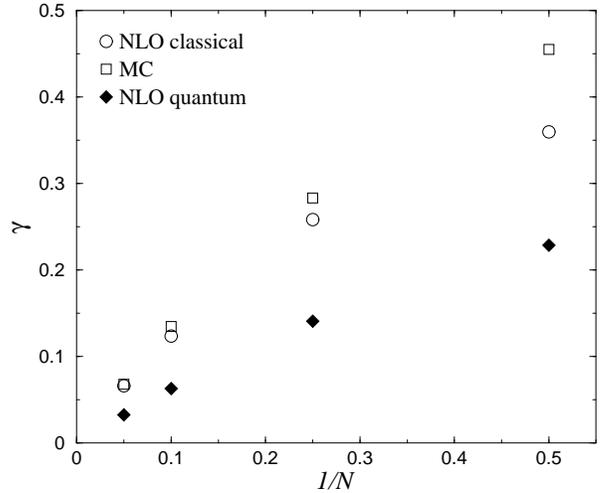,height=6.8cm}
\end{center}
\caption{Nonequilibrium damping rates extracted from $F(t,0;p=0)$ shown in
Fig.~\ref{fig1} as a function of $1/N$. Open symbols represent exact and
NLO classical evolution. One observes a rapid convergence of the $1/N$
expansion at NLO to the exact MC result. 
The quantum results are shown with full symbols. In the quantum theory
the damping rate is reduced compared to the classical theory. 
} 
\label{fig2}
\end{figure}

{\em Convergence of NLO and Monte Carlo results.} 
In Fig.~\ref{fig1} we present the two-point function $F(t,0;p=0)$ in the
classical field theory limit for three values of $N$. All other parameters
are kept constant. The figure compares the time evolution using the $1/N$
expansion of the $2PI$ effective action to NLO and the Monte Carlo
calculation that includes all orders in $1/N$.  One observes that the
approximate time evolution of the correlation function shows a rather good
agreement with the exact result even for small values of $N$ (note that
the effective four-point coupling is strong, $\lambda/6N=2.5 m_R^2$ for
$N=2$). For $N=20$ the exact and NLO evolution can hardly be
distinguished. A very sensitive quantity to compare is the damping rate
$\gamma$, which is obtained from an exponential fit to the envelope of
$F(t,0;p=0)$.  The systematic convergence of the NLO and the Monte Carlo
result as a function of $1/N$ can be observed in Fig.~\ref{fig2}.  The
quantitatively accurate description of far from equilibrium processes
within the NLO approximation of the $2PI$ effective action is manifest.

{\em Classical behavior of nonequilibrium quantum fields.}
In Fig.~\ref{fig2} we also show the damping rate from the quantum
evolution, using the same initial conditions and parameters. We observe
that the damping in the quantum theory differs and, in particular, is
reduced compared to the classical result.  In the limit $N\to \infty$
damping of the unequal-time correlation function $F(t,0;p)$ goes to zero
since the nonlocal part of the self energies
(\ref{sigmaF})--(\ref{sigmarho}) vanishes and scattering is absent. In
this limit there is no difference between evolution in a quantum and
classical statistical field theory.

\begin{figure}
\begin{center}
\epsfig{file=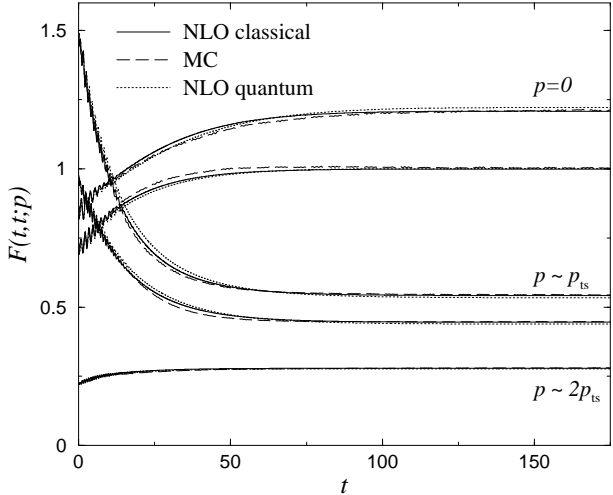,height=6.8cm}
\end{center}
\caption{Nonequilibrium evolution of the equal-time two-point function
$F(t,t;p)$ for $N=10$ for various momenta $p$. 
One observes a good agreement between the exact MC (dashed) and
the NLO classical result (full). The quantum evolution is shown with
dotted lines. The initial particle density is six times as high as
in Figs.\ \ref{fig1},\ref{fig2}. At these high densities, the difference
between quantum and classical evolution is small.
}\label{fig3}
\end{figure}

For finite $N$ scattering is present and  quantum and classical
evolution differ in general. However, the classical field approximation
may be expected to become a reliable description for the quantum theory if
the number of field quanta in each field mode is sufficiently high. 
We observe that increasing the initial particle number density leads
to a convergence of quantum and classical time evolution at not too late
times.
In Fig.~\ref{fig3} we present the time evolution of the equal-time
correlation function $F(t,t;p)$ for several momenta $p$ and $N=10$. Here
the particle density $\int \frac{dp}{2\pi} n_0(p)/M=1.2$ is six
times as high as in Figs.~\ref{fig1},\ref{fig2} and, in contrast to the
latter,
quantum and classical evolution at NLO follow each other rather
closely. For an estimate of the NLO truncation error we also give the MC
result for $N=10$ showing a quantitative agreement with the classical NLO
evolution both at early and later times.

{\em Quantum versus classical equilibration.}  
{}From Fig.~\ref{fig3} one observes that the initially highly occupied 
``tsunami'' modes ($p_{\rm ts}/m_R = 2.5$) ``decay'' as time proceeds
and the
low momentum modes become more and more populated. At late times
the classical theory \cite{Aarts:2000mg,Aarts:2001wi}
and the quantum theory \cite{Berges:2000ur,Berges:2001fi} approach their 
respective thermal equilibrium distribution. Since classical and quantum 
thermal equilibrium are distinct the classical and quantum time evolutions
have to deviate at sufficiently late times, irrespective of the
initial particle number density per mode.
Differences in the particle number distribution can be conveniently
discussed using the inverse slope parameter
$T(t,p) \equiv - n(t,\epsilon_p) \mbox{$[n(t,\epsilon_p)+1]$} 
(dn/d\epsilon)^{-1}$
for a given time-evolving particle number distribution $n(t,\epsilon_p)$
and dispersion
relation $\epsilon_p(t)$ \cite{Berges:2001fi}.
Following Ref.\ \cite{Aarts:2001qa} we define the effective
particle number as     
$
n(t,\epsilon_p)+\frac{1}{2}
\equiv [F(t,t';p)\, \partial_{t}\partial_{t'} 
F(t,t';p) ]^{1/2}|_{t=t'}
$ 
and mode energy by
$
\epsilon_p(t) \equiv [\partial_{t}\partial_{t'} 
F(t,t';p)/F(t,t';p)]^{1/2}|_{t=t'}
$,
which coincide with the usual free-field definition for $\lambda \to 0$. 
For a Bose-Einstein distributed particle number the parameter $T(t,p)$
corresponds to the (momentum independent) temperature $T(t,p)=T_{\rm eq}$.  
In the classical limit the inverse slope $T(t,p)$ as
defined above remains momentum dependent.

\begin{figure}
\begin{center}
\epsfig{file=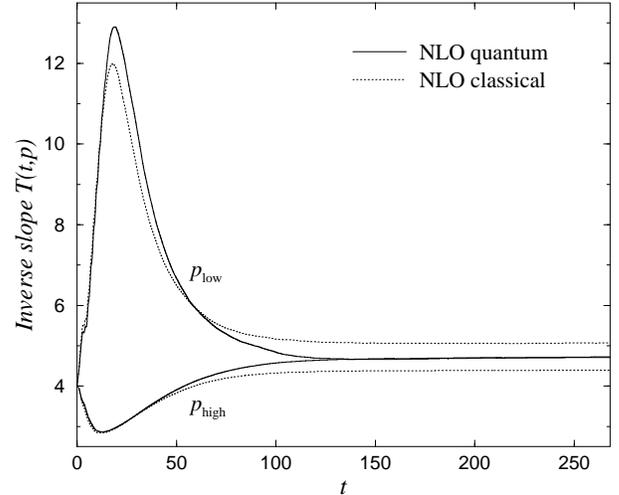,height=6.8cm}
\end{center}
\caption{Time dependence of the inverse slope $T(t,p)$, as defined in the
text. When quantum thermal equilibrium is approached, all modes get equal
inverse slope. In contrast, for classical thermal equilibrium the inverse
slope is momentum dependent with $T(p_{\rm low}) > T(p_{\rm high})$. 
}
\label{fig4}
\end{figure}

In Fig.~\ref{fig4} we plot the function
$T(t,p)$ for $p_{\rm low}\simeq 0$ and $p_{\rm high} \simeq 2 p_{\rm ts}$. 
Initially one observes a very different behavior of $T(t,p)$ for the low
and high momentum modes, indicating that the system is far from
equilibrium. Note that classical and quantum evolution agree very well for
sufficiently high initial particle number density.
However, at later times the difference between quantum and classical
evolution becomes visible. The quantum evolution approaches
quantum thermal equilibrium with a momentum independent inverse slope 
$T=4.7m_R$. 
In contrast, in the classical limit the slope parameter remains momentum
dependent and the system relaxes towards classical thermal equilibrium 
\cite{ClTh}. 

We thank W.\ Wetzel for continuous support in computational resources. 
This work was supported by the TMR network {\em Finite
Temperature Phase Transitions in Particle Physics}, EU contract no.\
FMRX-CT97-0122.

\end{document}